%% file: main.tex
\documentclass[journal]{IEEEtran}
\usepackage{url}
\usepackage[utf8]{inputenc}
\usepackage{cite}
\usepackage{amsmath,amssymb,amsfonts}
\usepackage{algorithm,algpseudocode}
\usepackage{graphicx}
\usepackage{textcomp}
\usepackage{xcolor}
\usepackage{caption}
\usepackage{subcaption}
\usepackage{multicol}
\usepackage{multirow}
\usepackage{listings}
\usepackage{adjustbox}

\def\BibTeX{{\rm B\kern-.05em{\sc i\kern-.025em b}\kern-.08em
    T\kern-.1667em\lower.7ex\hbox{E}\kern-.125emX}}

\begin{document}

\title{Reverse Engineering Word-Level Models from Look-Up Table Netlists}
\date{November 11 2022}
\author{\IEEEauthorblockN{Ram Venkat Narayanan, Aparajithan Nathamuni Venkatesan, \\ 
Kishore Pula, Sundarakumar Muthukumaran and Ranga Vemuri} \\
\IEEEauthorblockA{\textit{Digital Design Environments Lab, ECE Department} \\
\textit{University of Cincinnati, Cincinnati, Ohio, USA}\\
narayart@mail.uc.edu, nathaman@mail.uc.edu, \\ 
spulake@mail.uc.edu, muthuksr@mail.uc.edu , vemurir@mail.uc.edu}\\}
\maketitle

\input{Abstract}
\input{Introduction}
\input{Background}
\input{Carry_Chain_Analysis}

\input{Overall_Tool}

\input{Results}

\input{Future_work}

\input{Acknowledgment}
\nocite{*}
\bibliographystyle{ieeetr}
\bibliography{main}

\end{document}

%% file: Abstract.tex

\textbf{Reverse engineering of FPGA designs from bitstreams to RTL models aids in understanding the high level functionality of the design and for validating and reconstructing legacy designs. Fast carry-chains are commonly used in synthesis of operators in FPGA designs. We propose a method to detect word-level structures by analyzing these carry-chains in LUT (Look-Up Table) level netlists. We also present methods to adapt existing techniques to identify combinational operations and sequential modules in ASIC netlists to LUT netlists. All developed and adapted techniques are consolidated into an integrated tool-chain to aid in reverse engineering of word-level designs from LUT-level netlists. When evaluated on a set of real-world designs, the tool-chain infers 34\% to 100\% of the elements in the netlist to be part of a known word-level operation or a known sequential module.}

%% file: Introduction.tex
\section{Introduction} \label{Intro}

FPGA reverse engineering plays an important role in understanding potential security concerns. Reverse engineering of RTL models from logic-level netlists is also helpful in understanding adversarial designs for which an RTL description of the design is unavailable. However, the heterogeneous functionalities of Configurable Logic Blocks (CLBs) in today's FPGAs increase the complexity of reverse engineering to detect word-level structures implemented in the FPGA fabric. 

There are three stages to reverse engineering an FPGA design - \textit{bitstream extraction}, \textit{netlist extraction} and \textit{specification discovery} \cite{surveyFPGA}. \textit{Bitstream extraction} is the process of recovering the bitstream file that is used to configure an FPGA. The bitstream file can be intercepted during the boot-up process of an FPGA \cite{FPGASecurity}. \textit{Netlist extraction} is the process of recovering the LUT-level netlist from the extracted bitstream. FPGAs are made up of a series of CLBs. Depending on the type of FPGA, the CLB can contain different logic primitives. Modern FPGAs have several primitives such as Look-Up-Tables (LUT), storage elements such as latches and flip-flops, shift registers, RAM blocks, multiplexers, and carry blocks. Specification discovery is the process of extracting high-level functional modules from gate-level or LUT-level netlists. The extracted functional modules can be used to generate a RTL model in VHDL or Verilog. In this paper, we assume that the bitstream extraction and netlist extraction are completed and focus on the problem of specification extraction from flate netlists of LUTs, Carry modules and flip-flops. There are excellent tools such as Project X-ray \cite{ProjectXray} for LUT-level netlist reverse engineering.

For the purpose of reverse engineering, we assume a flattened LUT-level netlist. Therefore, we lose the original modular hierarchy of the netlist. In addition, word-level structures in the design might be cross-optimized with other components in the netlist. We also assume that we do not possess any information about the names of the resources and nets in the flattened netlist. The primary goal of this paper is to provide an automated tool that can analyze the flattened LUT-level netlist and generate a high-level representation for this netlist. A high-level representation of the netlist not only abstracts the netlist description to the word-level but also provides a modular hierarchy to help understand the netlist. Our tool-chain analyzes carry-chains in FPGA netlists to identify word-level operations and ALUs. In addition, the tool-chain adapts state-of-the-art ASIC reverse engineering techniques to identify sequential components and other word-level combinational operations in LUT-level netlists. 

This paper is organized as follows. Section \ref{SOTA} discusses the state-of-the-art ASIC and FPGA gate-level netlist reverse engineering techniques. Section \ref{WordLevelCarryChain} discusses our new research in how carry-chains are analyzed to identify word-level operations and ALUs. Section \ref{Reverse engineering framework for FPGA designs} highlights the state-of-the-art techniques adapted by the tool and shows how the different techniques work in tandem to obtain a top-level RTL model. The tool is evaluated on a set of real-world designs and the results are presented in Section \ref{Results}.


%% file: Background.tex
\section{Related Work} \label{SOTA}
\subsection{RTL or word-level models for ASIC netlists}
Subramanyan et al.
\cite{subramanyan2013reverse} focused on the detection of commonly used datapath components such as adders, multiplexers, counters, registers, and RAM modules in ASIC gate-level netlists using structural and functional analyses. A gate coverage metric was used to quantify their results. The paper shows that the components detected varied from 45\% to 94\% in real-world benchmarks. Gascon et al. \cite{templateCircuit} proposed a method to identify commonly used word-level modules by means of Permutation-Independent Equivalence Checking (PIEC). The techniques were evaluated successfully on a set of reverse engineering benchmarks from ISCAS'85 and academic implementations of ALUs. Meade et al. \cite{NetA} provided a toolset for segregating control logic and datapath logic in ASIC gate-level netlists. They analyzed the structure of the fan-in for each gate in the netlist and grouped elements based on similarity. They also provided a way to obtain state registers from the control logic and further construct FSMs and describe them using a high-level description. However, the techniques described in these papers were tested only on gate-level netlists of ASICs. 

\subsection{FPGA bitstream to netlist reverse engineering}
Benz et al. \cite{BIL} present a tool-chain for partial conversion of a FPGA bitstream to LUT level netlist. The tool implements a correlation technique by using the XDLRC file available in Xilinx FPGAs. The bitstream reversal tool utilizes this XDLRC to map the XDL components to the bits in the bitstream and produces a netlist in XDL file format. The tool is capable of extracting only a fraction of the netlist from the bitstream.

Zhang et al. \cite{Toolchain} present a tool-chain to obtain an RTL description in Verilog from a FPGA bitstream. The tool consists of Library Generator (LG), Bitstream Reversal Tool (BRT), and a Netlist Reversal Tool (NRT). The LG maps the bitstream to logic primitives in the configurable logic blocks. The NRT includes a procedure in which a reverse BFS from the primary output and ending with the primary inputs extracted the clusters in the netlist. Note that the BFS performed from primary output to the primary input does not recover the module boundaries or reveal information about the actual hierarchy of the design or the word-level groupings of elements.

\subsection{Netlist to RTL description for FPGAs}
Albartus et al. \cite{DANA} provided techniques to identify register groupings in both FPGA and ASIC gate-level netlists. They used control signals, neighborhood information, and other structural information to obtain highly accurate groupings. Fyrbiak et al. \cite{GraphSim} proposed a way to detect trojans in FPGA and ASIC designs by comparing subcircuits with library modules. The library modules were un-tampered circuits and a graph similarity measure was used to compare with the target subcircuit to check for Trojans. However, these techniques do not directly provide insight into the different operators and other datapath components in the circuit. 

Although there is significant research done in the field of gate-level ASIC netlist reverse engineering, only a few techniques address FPGA designs. In FPGAs, LUTs, carry modules and other combinational modules absorb considerable amount of logic. Methods that inspect the subcircuit structure containing gates in ASIC netlists cannot be immediately applied to FPGA netlists. Most of the current research on FPGA netlist reverse engineering uses structural analysis and graph comparison techniques to identify word-level modules and flip-flop groupings in a netlist. Current research also converts the LUTs, carry blocks and other combinational modules to a generic representation such as the AND-OR-INV logic \cite{GraphSim}. However, some of the structural information is lost by converting the design to such a generic representation. 

The tool-chain presented in this paper identifies datapath components and word-level modules by using structural and functional analyses. The tool-chain also analyzes fast carry-chains in modern FPGA netlists to identify word-level operations. Even though the tool-chain described in this paper is specific to current Xilinx FPGA architectures, the techniques can be extended to include other FPGA families by minimal modifications to the algorithms. Along with the identification of components, the tool-chain also produces a high-level representation in Verilog. 

%% file: Carry_Chain_Analysis.tex
\section{Extracting Word-level Operators Using Carry-Chains} \label{WordLevelCarryChain}

Section \ref{carry-chains} provides necessary background on carry-chains in Xilinx FPGAs. Section \ref{cca} discusses how carry-chain analysis is used to detect arithmetic and comparison operations. Section \ref{ccaALU} shows how carry-chain analysis is extended to detect ALUs. 

\subsection{Carry-chains} \label{carry-chains}
To significantly improve performance in common arithmetic operations, modern FPGAs are designed with low-level primitives that can perform specific tasks in a highly optimized manner. One such primitive is the carry module. Hauck et al. \cite{carry_efficiency} shows how significant performance can be achieved by using carry-chains in FPGAs. Therefore, using the carry module is an integral part of the synthesis of operations in both Xilinx and Altera FPGAs\footnote{Now AMD and Intel respectively}. 

\begin{figure} [h!]
    \centering
    \includegraphics[width=0.55\linewidth]{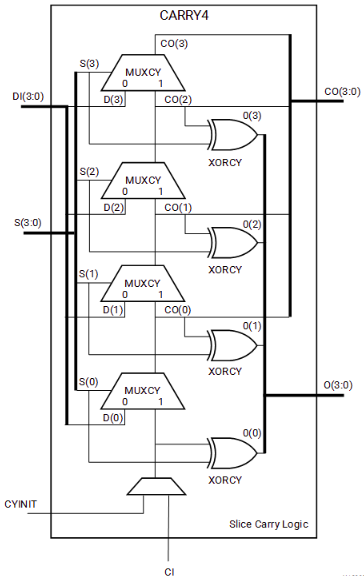}
    \caption{CARRY4 element in Xilinx 7-series FPGAs \cite{CARRY4documentation}.}
    \label{carry_element}
\end{figure}

The carry module in Xilinx FPGAs shown in Fig. \ref{carry_element} consists of MUXCY multiplexers and XORCY xor gates \cite{CARRY4documentation}. The MUXCY component selects the \textit{carry-in} for each bit. The XORCY component performs the arithmetic operation. The select line for the multiplexer MUXCY is provided through the input S and one of the data inputs to the multiplexer is provided through the input DI. The carry module performs a specific function based on the inputs provided on the S, DI, CYINIT and CI pins. Multiple carry modules are cascaded to perform wider arithmetic logic. By analyzing the function performed by the logic in the transitive fan-in of each input pin in the carry-chain, we can identify the operation performed by the chain. 

In Xilinx 7-series architecture, carry-chains are used to implement arithmetic and comparison operations. Bitwise operations are implemented directly using LUTs. Arithmetic operations are word-level operations that produce a multi-bit word as the output. On the contrary, comparison operations produce a single bit as the output. Therefore, by analyzing the output pin in the carry chain, we can differentiate between pure arithmetic and comparison operations. 

\begin{figure}[!htb]
     \centering
     \includegraphics[width=\linewidth]{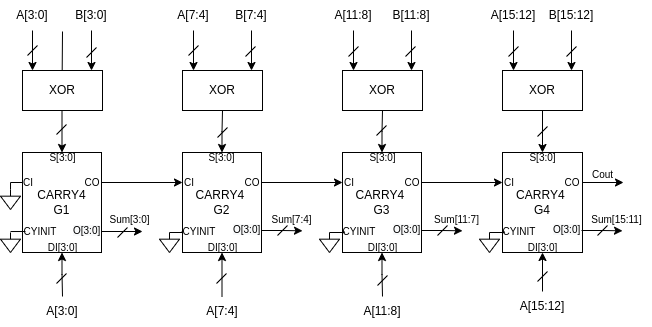}
     \caption{I/O configuration for a 16-bit adder implemented using a carry-chain} 
     \label{Adder carry chain}
\end{figure}

As an example, Figure \ref{Adder carry chain} shows the logic connected to the input and output pins for a 16-bit adder. The two sum operands are XORed and provided as input to the S pin. The DI pin receives input from either of the operands. The DI pin is the data input '0' of the MUXCY. The select line of the MUXCY is the result of the XOR operation between the two input operands. The select line is low if and only if both the operands are the same. This is why the DI pin can receive either of the operands. The CYINIT pin receives a '0' since addition is performed. For a subtraction, one of the operands is inverted and the CYINIT pin receives a '1' in order to complete the two's complement negation. The output word is taken from the O pins of the carry-chain. 

\subsection{Carry-Chain Analysis} \label{cca}
Carry elements are utilized for implementing arithmetic and comparison operations. Carry elements are considered as the starting point in detecting these operators. Once the different carry-chains in the netlist are identified, we perform the analysis on the input pins of each CARRY4 gate in the carry-chain. A reverse Breadth-First Search (BFS) is performed at each input pin of the CARRY4 element in the carry-chain to identify the elements in the transitive fan-in of each pin. 
BFS traversal stops at a \emph{potential} module boundary. 

Algorithm \ref{Detect carry chain} shows the procedure to detect carry-chains in LUT-level netlists. The input to the procedure is the list of all carry elements in the netlist. For each carry element in the netlist, we first check the connection to the CI pin. If the pin is connected to ground, the carry element is the start of the carry-chain. Lines 5-10 are executed when the start of the chain is identified. In line 6, the starting carry element is added to the carry-chain. Setting the starting carry element as the current carry element, we look at the CO[3] pin of the element. If there is another carry element connected to the current element, we append it to the chain and update the new carry element with the current carry element. We repeat this process until we cannot find another carry element connected to the CO[3] pin. This helps us find a single carry-chain in the netlist. In line 10, we append the chain to a list of carry-chains. When this is run across all the carry elements in the netlist, we can identify all the carry-chains. The function returns the list of all carry-chains.

\begin{algorithm}[h!]
  \small
  \caption{Detection of carry-chains}\label{Detect carry chain}
  \begin{algorithmic}[1]
    \Procedure{detectCarryChains}{$carryGates$}
    \State \texttt{$carryChains \gets [ ]$}
    \For{$carryGate$ in $carryGates$}
        \If{CI pin of $carryGate$ is 0}
            \State $currentChain \gets [carryGate]$ 
            \While{CO[3] of $carryGate$ is linked to CARRY4}
                \State $carryGate \gets carryGate$ linked to CO[3]
                \State $currentChain.append(carryGate)$
            \EndWhile
        \State $carryChains.append(currentChain)$
        \EndIf
    \EndFor
    \State \textbf{return} $carryChains$
    \EndProcedure
    \end{algorithmic}
\end{algorithm}

\subsubsection{Detection of Pure Operations}
A potential module boundary is delineated by flip-flops, RAM blocks, primary inputs and other CARRY4 elements which are potential input sources (operand) to an operator. A logic function is obtained for each pin in terms of the nets at the modular boundary by using all the elements in the transitive fan-in of the corresponding pin. In pure operations, a carry-chain implements only a single operation. Therefore, in pure operations, the fan-in function is same at all the S pins and the fan-in function is same at all the DI pins. A permutation-independent form for the fan-in function at the S and DI pins of a CARRY element is obtained using the fast Boolean matching technique presented by Huang et al. \cite{booleanmatching}. By comparing the permutation-independent form of the logic functions with a library, pure operations can be detected. For each operation, the library contains the corresponding permutation-independent fan-in functions at the S and DI pins of the CARRY4. For example, the adder entry in the library contains the permutation-independent form of the XOR function at the S pin and the permutation-independent form of the BUFFER function\footnote{Here, BUFFER function refers to the function Y = A} at the DI pin. 

Algorithm \ref{Standalone operation detection} shows the procedure to detect pure operations using carry-chains. The function receives the carry-chain as the input and returns the elements that are part of the carry-chain operation and the identified operation as the output. If the operation is unknown, \textit{None} is returned. At first, for each S and DI pin in each carry element in the carry-chain, we perform a reverse BFS traversal and identify the elements in the fan-in. All the elements in the traversal are added to the module. This is shown in line 7. If the number of inputs in the fan-in is less than six, it signifies a potential pure operation that can be detected. This is shown in line 8. The number was chosen after analyzing the number of inputs in commonly used pure carry-chain-based operations. For pure carry-chain-based operations such as addition, subtraction and comparison, the number of inputs in the fan-in were identified to be less than six. Using the elements in the fan-in, we first obtain a fan-in equation and then identify a permutation-independent canonical form for the equation. This is shown in lines 9-10. We examine the output nets at the O and CO pins of the carry-chain to identify the module boundary for the operation. If the output nets are connected to a flip-flops, the flip-flops are added to the list of elements in the module as shown in line 15. In this case, the output word is mapped to the output of the flip-flops. However, if the output nets are not connected to flip-flops, the output word is mapped the output pins of the carry-chain. The operation is identified by comparing the S and DI fan-in functions of the candidate carry-chain and a library of S and DI fan-in functions. 

\begin{algorithm}[h!]
  \caption{Detection of pure operations using carry-chains} \label{Standalone operation detection}
  \begin{algorithmic}[1]
    \Procedure{identifyOperation}{$carryChain$}
    \State {$operation \gets "unknown"$}
    \State $moduleGates \gets [\ ]$
    \For{$carryGate$ in $carryChain$}
        \For {$pin$ in $carryGate$} \Comment{input pins}
            \State $fanIn \gets$ extractFanIn($pin$)
            \State $moduleGates.add(fanInGates)$
            \If{no. of inputs is $<$ 6} 
                \State $pinEqn \gets$ getEquation($fanIn$,$pin$) 
                \State $pinEqn \gets$ getCanonical($pinEqn$)
            \EndIf
        \EndFor
        \For {$pin$ in $carryGate$} \Comment{output pins}
            \If{output pin is connected to a flip-flop}
                \State $moduleGates .append($flip-flop$)$
            \EndIf
            
        \EndFor
    \EndFor
    \If{function of S, DI pins of $carryGate$ in library}
        \State $operation \gets$ matched library function
    \EndIf
    \State \textbf{return} $moduleGates, operation$
    \EndProcedure
    \end{algorithmic}
\end{algorithm}

\subsubsection{Detection of Cross-Optimized Operations}
Since a standard structure is found in pure operations, it is straightforward to identify these operations. However, we need to devise a different approach to identify the functionality of a carry-chain that performs more than one operation.  An \textit{Add\_Sub} operation performs either an addition or a subtraction based on a specific condition. In ASIC netlists, multiplexers are used at the input or output of an \textit{Add\_Sub} operation to choose between an addition or a subtraction. Since LUTs absorb the multiplexer logic, an \textit{Add\_Sub} operation synthesized using an FPGA might not implement separate addition and subtraction modules. In order to identify carry-chains that perform the \textit{Add\_Sub} operation that are dependent on a select line, we first identify the select line in the transitive fan-in of the input pins. 

We identify select lines by analyzing the input nets in the fan-in of S pin for each carry element in the carry-chain. A net connected to all S pins of a carry-chain is classified as a candidate select line. The carry-chain implements an operation based on the signal value on the select line. For each input combination of the select line, we identify the fan-in function of the carry elements in the carry-chain. By comparing the permutation-independent form of the fan-in functions with a library, the operation can be detected. Thereby, we identify the operation performed by the carry-chain for each combination of the select line.

\subsubsection{RTL representation}
In FPGA netlists, the input and output pin ordering of carry modules can be used to correctly identify the input and output word. Using the ordering information, the operations can be extracted to a higher level of abstraction. Input pins S[0] to S[3] and DI[0] to DI[3] of the first CARRY4 in the chain corresponds to the first four bits of the input word. The S[0] to S[3] and DI[0] to DI[3] of the cascaded CARRY4s in the chain correspond to the respective bits of the input word. 

Pins S[0:3] and DI[0:3] have the same connectivity for all carry-chains. The first carry element is indicated by the CI pin being set to '0'. The last carry element is indicated by the CO pin not connected to another carry element. The input and output pin orderings of carry elements aid in understanding the order of bits in the words obtained in the previous steps. The first and second input words are obtained by comparing the inputs in the fan-in of the S and DI input pins of carry elements with a reference library. For each operation, this reference library contains the corresponding sequence in which the input operands appear at the fan-in of the S and DI pin. For example, the adder entry in the library contains ((1,A)) as the input to the fan-in cone of DI pin and contains ((1,A),(2,B)) as the inputs to the fan-in cone of the S pin. The numbers represent the sequence and the letter represents the corresponding input word.

The bits of the output word are simply the output nets of the carry-chain. However, this technique only works when the ordering of inputs in the transitive fan-in of each pin is the same. If the order of the inputs in the transitive fan-in is changed, we need to first differentiate between the input operands. Then, the ordering can be identified using the carry-chain. Flip-flops are the most common source of inputs. Flip-flop grouping techniques can be used to differentiate between different input words (see Section \ref{Detection of Sequential Modules}). 

\subsection{Detection of ALUs} \label{ccaALU}
Inspecting the fan-in cone at the input of each pin of a carry element is insufficient to detect words when the logic contained in the fan-in-cones is complex. Therefore, a different approach is required to identify complex combinational logic such as the ALU. To this end, potential ALUs are first identified by obtaining the register boundaries at the input and output of the carry-chain. For identifying the functionality of the ALU, we use the Quantified Boolean Formula (QBF) solver approach described by Gascon et al. \cite{WordRev}. However, we adapt it to FPGA netlists.

To identify the ALU boundaries, we first apply a reverse BFS to obtain the input word. To obtain the output word, we perform forward BFS starting at the CARRY elements and stopping at a potential module boundary (flip-flop, RAM module and primary output). All the elements encountered in this traversal might not be part of the same carry-chain-based operation. Therefore, a reverse BFS technique starting at the carry elements and stopping at the potential module boundaries is used to identify the inputs for these elements. By doing this set of traversals, a register-to-register boundary that contains carry-chains interspersed with other combinational logic elements is obtained. A typical ALU receives two input words and produces an output word based on a condition. This condition is usually referred to as the \textit{opcode}. If the register-to-register boundary obtained matches the standard structure of an ALU, it is marked as a potential ALU.



\begin{figure}[!htb]
     \centering
     \includegraphics[width=0.76\linewidth]{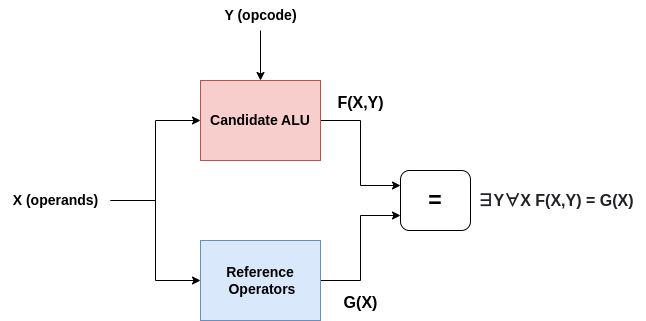}
     \caption{QBF formulation} 
     \label{QBF_ALU}
\end{figure}

Figure \ref{QBF_ALU} shows the QBF formulation setup to identify a given ALU's functions.
Once the ALU boundaries are identified, we identify the functions of the candidate ALUs by comparing it with a library of known components. In this library, we contain a Verilog description of a list of commonly used operations that include arithmetic, comparison, logical, bitwise and shifting operations. Each operation in this library is designed with a generic width, say N. The width N is updated to match the candidate ALU width during the comparison. To compare with the reference operator, the candidate ALU is represented using a Verilog description where the LUTs, carry modules and multiplexers in the module using a simplified form of their Boolean function. \textit{Operand A} and \textit{Operand B} of the ALU are provided as inputs to the reference operator circuit and the candidate ALU. The candidate module receives the opcode as side inputs. The equivalence function is used to form the Miter circuit to test for functional equivalence between the outputs of the two circuits. The QBF solver checks whether an \textit{opcode} exists such that the two circuits are equivalent for all input combinations. By testing the candidate ALU with various reference operators, we can identify the set of functions performed by the ALU and the corresponding \textit{opcode} values.

%% file: Overall_Tool.tex
\section{Reverse Engineering Tool-Chain for FPGA Designs} \label{Reverse engineering framework for FPGA designs}

The proposed tool-chain uses the carry-chain analysis technique along with existing state-of-the-art techniques to identify word-level modules in ASIC netlists, which we adapted to LUT netlists. To detect bitwise operations, counters, and shift registers, we use the techniques proposed by Subramanyan et al. \cite{subramanyan2013reverse}. Section \ref{kca} shows how K-cut analysis is used on FPGA designs to identify bitwise operations. Section \ref{Detection of Sequential Modules} shows how grouping and graph isomorphism techniques are used to identify sequential modules.

\subsection{K-Cut Analysis} \label{kca}
Word-level operators are usually synthesized into bit-sliced designs as it is the most efficient way to perform word-level operations. The functions of these bit-slices are usually derived using a collection of logic gates. Since LUT-level netlists are synthesized using LUTs, we analyze the INIT values of the LUTs to identify the logic functions. Therefore, to identify the logic used in a bit-slice, we need to determine the set of LUTs that perform the logic. Cut enumeration is used to detect bit-slices. In a netlist, a cut of a net \textit{N} refers to a set of nodes that are present in the transitive fan-in of \textit{N} that can definitively assign a logic value to net \textit{N}. When the number of nets that are used to define \textit{N} is less than or equal to \textit{k}, it is called a \textit{k-cut}. Most of the operators in FPGA designs are synthesized using carry-chains. However, pure bitwise operations are not synthesized using carry-chains. Therefore, to improve the effectiveness of detecting word-level operations, we include this analysis. 

For an n-bit bitwise operation with \textit{Out} as its output and \textit{A} and \textit{B} as its inputs, \textit{Out[i]} is only dependent on \textit{A[i]} and \textit{B[i]}. Thus, it is not possible to identify a word operation only using knowledge about the functionality of the bit-slices. To identify word-level bitwise operations, we analyze the neighborhood information along with the functions of each bit-slice. In real-world designs, these pure bitwise operations usually receive their inputs from a group of flip-flops and the outputs are connected to a group of flip-flops. To differentiate between flip-flops that are inputs and outputs to the bitwise operation from the other flip-flops in the netlist, the register stage identification algorithm provided in DANA by Albartus et al. \cite{DANA} is used.

\subsection{Detection of Sequential Modules} \label{Detection of Sequential Modules}
When reverse engineering FPGAs, it is essential to apply techniques to detect sequential components formed by flip-flops. We use techniques described in \cite{subramanyan2013reverse} for identifying sequential components. Techniques described in \cite{DANA} can also be used to identify sequential modules. Counters, shift registers, registers, and RAM modules are some of the commonly used sequential components. Xilinx's 7-series FPGAs use embedded memory resources in their FPGAs such as Block RAMs to implement designs that require RAM modules \cite{Memorydocumentation}. Thus, the goal is to detect counters, shift registers, and registers. 

To ensure the synchronization of flip-flops in a register, these flip-flops usually have the same clock and control signals. Thus, grouping flip-flops in a netlist based on control signals provides us with potential registers in the netlist. Even though the idea behind this technique is very simple, it is an effective way to detect module boundaries in a netlist.

Due to the nature of the operation of counter and shift register circuits, the data flow between the flip-flops in these sequential components follows a standard structure. By analyzing the data flow between the flip-flops in a register grouping, we can identify counters and shift registers by comparing their Flip-Flop Connectivity Graphs (FFCGs)\footnote{This is similar to the Latch Connection Graph (LCG) mentioned in \cite{subramanyan2013reverse} and Data Flow Graph (DFG) mentioned in \cite{DANA}}. A flip-flop connectivity graph is a graph data structure in which the nodes are flip-flops and the edges are connections between the flip-flops. For constructing the FFCG, we replace any combinational logic element (logic gates in ASICs) encountered in the traversal is replaced by a wire.

Once the different flip-flop groupings and sequential components in the circuit are identified, a modular boundary is outlined by identifying the gates that are part of the module. This is done by analyzing paths between each pair of flip-flops in a grouping. We perform forward BFS at the output of each flip-flop in the grouping and exit when we encounter another flip-flop. If the destination flip-flop obtained in each traversal is part of the same grouping, we add the combinational elements encountered in these paths as part of the module. If the destination flip-flop is not part of the same grouping, we disregard the combinational elements in this path as they are not part of this module. When this is done on all the paths, a complete modular boundary is obtained. 

Counters and shift registers are compared with a reference counter and shifter of the same size. If there is a perfect match in the mapping between all the input and output signals in the reference component and the inferred component, the component is extracted to a high-level description. If the structural mapping fails, we write a behavioral description for each flip-flop in the grouping using the logic expressions of the fan-in cones. The RTL for registers that are not classified as counters or shift registers are also written in a similar manner.


\subsection{Tool-Chain Workflow} \label{ToolWorkflow}

Sections \ref{WordLevelCarryChain} described how word-level operations can be detected using carry-chain analysis. Section \ref{kca} and Section \ref{Detection of Sequential Modules} show how ASIC reverse engineering techniques are adapted to FPGA designs. This section describes how the reverse engineering tool-chain integrates the techniques and produces a top-level module that instantiates the identified word-level modules. 

\begin{figure}[H]
     \centering
     \includegraphics[width=1\linewidth]{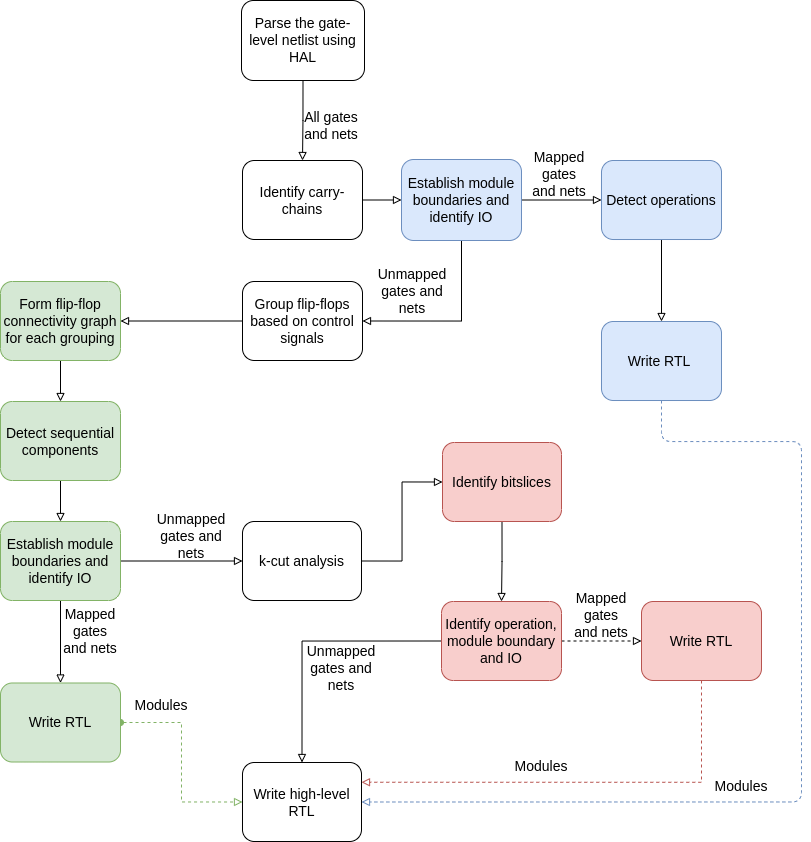}
     \caption{Workflow of the tool-chain}
     \label{tool_flow}
\end{figure}

The tool-chain integrates the techniques in a particular order and ultimately provides a high-level description of the gate-level netlist. Figure \ref{tool_flow} shows the order in which the techniques are integrated. HAL \cite{HAL} is used to parse the netlist. Then, the different carry-chains are identified in the netlist, and carry-chain-based operations are mapped as part of a module as shown in blue. The remaining unmapped gates and nets are passed on to the next stage where the flip-flops are identified and grouped based on control signals. Flip-flop connectivity graphs are formed on these groupings and sequential components are detected as shown in green. On the remaining unmapped nets and gates, k-cut analysis is performed and bit-slices are identified. Using bit-slices that represent the same functionality, bitwise operations are detected as shown in red. An RTL description is written for all the identified modules. They are instantiated in the top-level module to form a complete netlist. The obtained high-level RTL is then verified using JasperGold for equivalence checking with the source HDL to confirm functional validity. However, to ensure equivalence in sequential circuits, we reset all flip-flops in the source HDL that do not have an initialization.

%% file: Results.tex
\section{Experimental Results} \label{Results}
The techniques described in this paper were implemented using Python. HAL \cite{HAL} was used to perform fundamental operations on the netlist such as graph traversal and storing and retrieving data on design elements.  All the benchmark designs were first synthesized using Artix 7-series and Zync 7-series FPGAs using Xilinx Vivado. We set the \textit{flatten hierarchy} option to \textit{full}, while we kept the other default synthesis and optimization settings. We delete all the original net IDs that may indicate the signal name in the original model and generate unique random IDs for all nets. We use these netlists for our reverse engineering experiments. Yosys \cite{yosys} was used for the QBF SAT problem. Permutation-independent Boolean matching was performed using \textit{testnpn} \cite{booleanmatching} command in \textit{abc} \cite{abc}. All experiments were done on an AMD Ryzen 7 processor with 16GB RAM. 

A diverse set of real-world designs varying between arithmetic cores, DSP cores, and processors were chosen to perform this experiment. These benchmarks were taken from OpenCores \cite{OpenCores}. The number of carry modules, LUTs, flip-flops, and memory resources for each design are shown in Table \ref{BM list}. Section \ref{CCA_results} shows the number of carry-chains detected and the number of operations that were inferred in each benchmark. Section \ref{SC_results} shows the detection accuracy of sequential components. Section \ref{gatecoverage} shows the number of operations detected in each benchmark and the coverage achieved. Section \ref{exectime} discusses the execution time of the tool-chain. Section \ref{Discussion} discusses the overall results and provides ways to improve the gate coverage.

\begin{table}[!htb]
\centering
\caption{List of Benchmarks}
\label{BM list}
\begin{adjustbox}{width=0.48\textwidth}
\begin{tabular}{|c|l|c|c|c|c|c|}
\hline
Sl. No. & Benchmark              & LUTs & FFs  & Carrys & Mems & Total gates \\ \hline
1       & Simple 8-bit processor & 87   & 44   & 2      & 0    & 133         \\
2       & Hilbert transformer    & 232  & 525  & 60     & 0    & 817         \\
3       & UART                   & 528  & 360  & 4      & 6    & 898         \\
4       & Cordic polar2rect      & 739  & 688  & 175    & 0    & 1602        \\
5       & Cordic rect2polar      & 1021 & 1051 & 259    & 0    & 2331        \\
6       & OpenFPU                & 1591 & 760  & 115    & 0    & 2466        \\
7       & LXP32 processor        & 1999 & 1017 & 165    & 144  & 3325        \\
8       & M1 core                & 2724 & 1340 & 220    & 0    & 4284        \\
9       & Reed Solomon decoder   & 3700 & 2869 & 123    & 62   & 6754        \\
10      & AES                    & 3536 & 3968 & 0      & 172  & 7676        \\ 
11      & Canny edge detector    & 5262 & 4706 & 639    & 1981 & 12588       \\ \hline
\end{tabular}
\end{adjustbox}
\end{table}

\subsection{Carry-chain analysis} \label{CCA_results}

Table \ref{CCA_table} shows the number of gates and number of carry-chains detected along with the number of known operations detected for each benchmark. The column \textit{Detected Operations (\%)} shows the ratio of number of operations detected to the number of carry-chains in the benchmark. The column \textit{Converted to RTL (\%)} shows the percentage ratio of number of carry-chains that were extracted to RTL with respect to the number of carry-chains in the benchmark. The column \textit{Module Coverage \%} shows the percentage of elements in the netlist that were identified to be part of a module derived from the carry-chain. We include this percentage to highlight the number of gates that are part of a carry-chain-based operation in a netlist. The 128-bit AES core benchmark is not included because it did not contain any carry elements. We found that, on average, 54.19\% of the elements in these netlists were identified to be part of a carry-chain-based module. However, only 29.08\% of these modules were detected as a known operation. In the \textbf{Cordic core}, \textbf{Reed Solomon decoder} and \textbf{Canny edge detector} a large number of operations were detected. However, in the \textbf{Reed Solomon decoder} benchmark, detecting carry-chains was not enough to detect a significant portion of the netlist. Our tool successfully detected the ALU in the \textbf{Simple 8-bit microprocessor}. However, in the \textbf{M1 core} and \textbf{LXP32 processor}, we found that the register-to-register boundary did not match with the ALU structure. Thus, the operations detected were low in these designs . Since ALUs contain significant combinational logic, the coverage \% was also low in these designs. However, with the correct module boundary, we were able to identify the functionality of the ALU using the QBF solver technique.

\subsection{Detection accuracy of sequential modules} \label{SC_results}

Table \ref{SCA_table} shows the number of flip-flops and the number of sequential components detected in each benchmark. The columns \textit{Registers, Counters,} and \textit{Shifters} show the ratio of correctly identified components with respect to the total number of sequential components identified. The column \textit{Known Seq. coverage (\%)} shows the percentage of LUTs and flip-flops that were identified as part of a sequential module. All the flip-flops in the \textbf{128-bit AES core} received the same control signal. Therefore, no sequential components were identified. In the \textbf{Cordic core}, most of the flip-flops and LUTs were identified as part of a carry-chain based operation. On average, 27.04\% of the LUTS and flip-flops were identified to be part of a known sequential module.

\subsection{Gate coverage} \label{gatecoverage}

Table \ref{GC_COMP} shows the gate coverage for each benchmark. The \textit{Module (\%)} shows the gate coverage is a percentage of the number of gates identified to be part of a module with respect to the total number of gates. The \textit{Known component (\%)} shows the gate coverage as a percentage of the number of gates identified to be part of a module with respect to the total number of gates. Due to the high number of operations in the arithmetic and DSP cores, a high number of gates are identified as part of a known operation. In the \textbf{128-bit AES core} benchmark, 20 bitwise operations were identified. Due to the absence of carry-chains in this design, the inclusion of k-cut analysis was included to detect components in this design. On average, 56.12\% of the gates in the design were identified to be part of a known operation or sequential component. On average, 75.18\% of the gates in the design were identified as part of a module. 

\subsection{Execution time} \label{exectime}
The maximum time taken for execution of the tool-chain is around 5 minutes for the \textbf{128-bit AES core} benchmark. Since the design did not contain any carry-chains and no registers were grouped by shared control signals in this design, k-cut analysis was run on the entire netlist, resulting in a high execution time.

\subsection{Discussion} \label{Discussion}

The tool-chain produced good quality detection coverage on netlists of DSPs and arithmetic cores. By using state-of-the-art techniques to detect bit-slices and sequential components, we improved the coverage percentage on the UART and processor designs. Operations performed by the ALUs were detected by extending the carry chain detection algorithm assuming that we have the exact modular boundary of the ALU. By adapting select-line detection techniques to detect multiplexers and other grouping techniques, we can provide a comprehensive solution to detect ALUs. We were unable to identify the functionality of some carry-chain modules in these benchmarks as they did not match the structure of our library components. With the known module boundary, we can use the QBF-solving technique to detect the functionality of these modules.

\begin{table*}[!htb]
\centering
\caption{Operations detected using Carry-Chains}
\label{CCA_table}
\begin{adjustbox}{width=1\textwidth}
\begin{tabular}{clcccccccc|c|c|c|}
\hline
\multicolumn{1}{|l|}{Sl. No.} & \multicolumn{1}{l|}{Benchmark}              & \multicolumn{1}{l|}{Gates} & \multicolumn{1}{l|}{\begin{tabular}[c]{@{}l@{}}Carry-\\ Chains\end{tabular}} & \multicolumn{1}{l|}{Add} & \multicolumn{1}{l|}{Sub} & \multicolumn{1}{l|}{Comp} & \multicolumn{1}{l|}{ALU} & \multicolumn{1}{l|}{Add/Sub} & \begin{tabular}[c]{@{}l@{}}Detected \\ Operations (\%)\end{tabular} & \begin{tabular}[c]{@{}l@{}}Converted \\ to RTL (\%)\end{tabular} & \begin{tabular}[c]{@{}l@{}}Module\\ Coverage (\%)\end{tabular} & \begin{tabular}[c]{@{}l@{}}Known Operation\\ Coverage (\%)\end{tabular} \\ \hline
\multicolumn{1}{|l|}{1}       & \multicolumn{1}{l|}{Simple 8-bit processor} & \multicolumn{1}{l|}{133}   & \multicolumn{1}{l|}{1}                                                       & \multicolumn{1}{l|}{0}   & \multicolumn{1}{l|}{0}   & \multicolumn{1}{l|}{0}    & \multicolumn{1}{l|}{1}   & \multicolumn{1}{l|}{0}       & 100.00\%                                                            & 0.00\%                                                           & 22.79\%                                                        & 22.79\%                                                                 \\
\multicolumn{1}{|l|}{2}       & \multicolumn{1}{l|}{Hilbert transformer}    & \multicolumn{1}{l|}{817}   & \multicolumn{1}{l|}{15}                                                      & \multicolumn{1}{l|}{5}   & \multicolumn{1}{l|}{10}  & \multicolumn{1}{l|}{0}    & \multicolumn{1}{l|}{0}   & \multicolumn{1}{l|}{0}       & 100.00\%                                                            & 100.00\%                                                         & 80.85\%                                                        & 80.85\%                                                                 \\
\multicolumn{1}{|l|}{3}       & \multicolumn{1}{l|}{UART}                   & \multicolumn{1}{l|}{898}   & \multicolumn{1}{l|}{1}                                                       & \multicolumn{1}{l|}{0}   & \multicolumn{1}{l|}{0}   & \multicolumn{1}{l|}{0}    & \multicolumn{1}{l|}{0}   & \multicolumn{1}{l|}{0}       & 0.00\%                                                              & 0.00\%                                                           & 6.65\%                                                         & 0.00\%                                                                  \\
\multicolumn{1}{|l|}{4}       & \multicolumn{1}{l|}{Cordic polar2rect}      & \multicolumn{1}{l|}{1602}  & \multicolumn{1}{l|}{42}                                                      & \multicolumn{1}{l|}{0}   & \multicolumn{1}{l|}{0}   & \multicolumn{1}{l|}{0}    & \multicolumn{1}{l|}{0}   & \multicolumn{1}{l|}{20}      & 47.62\%                                                             & 47.62\%                                                          & 99.62\%                                                        & 48.12\%                                                                 \\
\multicolumn{1}{|l|}{5}       & \multicolumn{1}{l|}{Cordic rect2polar}      & \multicolumn{1}{l|}{2331}  & \multicolumn{1}{l|}{55}                                                      & \multicolumn{1}{l|}{0}   & \multicolumn{1}{l|}{0}   & \multicolumn{1}{l|}{0}    & \multicolumn{1}{l|}{0}   & \multicolumn{1}{l|}{15}      & 27.27\%                                                             & 27.27\%                                                          & 97.17\%                                                        & 40.71\%                                                                 \\
\multicolumn{1}{|l|}{6}       & \multicolumn{1}{l|}{OpenFPU}                & \multicolumn{1}{l|}{2466}  & \multicolumn{1}{l|}{17}                                                      & \multicolumn{1}{l|}{1}   & \multicolumn{1}{l|}{0}   & \multicolumn{1}{l|}{0}    & \multicolumn{1}{l|}{0}   & \multicolumn{1}{l|}{1}       & 11.76\%                                                             & 11.76\%                                                          & 41.19\%                                                        & 12.71\%                                                                 \\
\multicolumn{1}{|l|}{7}       & \multicolumn{1}{l|}{LXP32 processor}        & \multicolumn{1}{l|}{3325}  & \multicolumn{1}{l|}{40}                                                      & \multicolumn{1}{l|}{3}   & \multicolumn{1}{l|}{0}   & \multicolumn{1}{l|}{0}    & \multicolumn{1}{l|}{1}   & \multicolumn{1}{l|}{3}       & 15.00\%                                                             & 12.50\%                                                          & 52.43\%                                                        & 10.80\%                                                                 \\
\multicolumn{1}{|l|}{8}       & \multicolumn{1}{l|}{M1 core}                & \multicolumn{1}{l|}{4284}  & \multicolumn{1}{l|}{26}                                                      & \multicolumn{1}{l|}{1}   & \multicolumn{1}{l|}{1}   & \multicolumn{1}{l|}{4}    & \multicolumn{1}{l|}{1}   & \multicolumn{1}{l|}{2}       & 38.46\%                                                             & 23.08\%                                                          & 38.51\%                                                        & 19.22\%                                                                 \\
\multicolumn{1}{|l|}{9}       & \multicolumn{1}{l|}{Reed Solomon decoder}   & \multicolumn{1}{l|}{6754}  & \multicolumn{1}{l|}{41}                                                      & \multicolumn{1}{l|}{23}  & \multicolumn{1}{l|}{1}   & \multicolumn{1}{l|}{0}    & \multicolumn{1}{l|}{0}   & \multicolumn{1}{l|}{0}       & 58.54\%                                                             & 58.54\%                                                          & 22.11\%                                                        & 9.05\%                                                                  \\ 
\multicolumn{1}{|l|}{10}      & \multicolumn{1}{l|}{Canny edge detector}    & \multicolumn{1}{l|}{12588} & \multicolumn{1}{l|}{119}                                                     & \multicolumn{1}{l|}{75}  & \multicolumn{1}{l|}{10}  & \multicolumn{1}{l|}{2}    & \multicolumn{1}{l|}{0}   & \multicolumn{1}{l|}{0}       & 73.11\%                                                             & 73.11\%                                                          & 49.19\%                                                        & 40.22\%                                                                 \\ \hline
                              &                                             &                            &                                                                              &                          &                          &                           &                          &                              &                                                                     & Average                                                          & 54.19\%                                                        & 29.08\%                                                                 \\ \cline{11-13}

\end{tabular}
\end{adjustbox}
\end{table*}

\begin{table}[h!]
\centering
\caption{Detection of sequential components}
\label{SCA_table}
\begin{adjustbox}{width=0.48\textwidth}
\begin{tabular}{clccc|c|c|}
\hline
\multicolumn{1}{|l|}{Sl. No.} & \multicolumn{1}{l|}{Benchmark}              & \multicolumn{1}{l|}{FFs}  & \multicolumn{1}{l|}{Registers} & Counters & Shifters & \begin{tabular}[c]{@{}l@{}}Known Seq.\\ Coverage (\%)\end{tabular} \\ \hline
\multicolumn{1}{|l|}{1}       & \multicolumn{1}{l|}{Simple 8-bit processor} & \multicolumn{1}{l|}{44}   & \multicolumn{1}{l|}{4/4}       & 1/1      & 0/0      & 38.35\%                                                            \\
\multicolumn{1}{|l|}{2}       & \multicolumn{1}{l|}{UART}                   & \multicolumn{1}{l|}{360}  & \multicolumn{1}{l|}{8/9}       & 11/11/   & 1/1      & 54.23\%                                                            \\
\multicolumn{1}{|l|}{3}       & \multicolumn{1}{l|}{Hilbert transformer}    & \multicolumn{1}{l|}{525}  & \multicolumn{1}{l|}{4/4}       & 0/0      & 1/1      & 19.09\%                                                            \\
\multicolumn{1}{|l|}{4}       & \multicolumn{1}{l|}{Cordic polar2rect}      & \multicolumn{1}{l|}{688}  & \multicolumn{1}{l|}{0/1}       & 0/0      & 0/0      & 0.00\%                                                             \\
\multicolumn{1}{|l|}{5}       & \multicolumn{1}{l|}{OpenFPU}                & \multicolumn{1}{l|}{760}  & \multicolumn{1}{l|}{13/15}     & 0/0      & 0/0      & 50.69\%                                                            \\
\multicolumn{1}{|l|}{6}       & \multicolumn{1}{l|}{LXP32 processor}        & \multicolumn{1}{l|}{1017} & \multicolumn{1}{l|}{15/17}     & 2/3      & 0/0      & 23.37\%                                                            \\
\multicolumn{1}{|l|}{7}       & \multicolumn{1}{l|}{Cordic rect2polar}      & \multicolumn{1}{l|}{1051} & \multicolumn{1}{l|}{0/1}       & 0/0      & 0/0      & 0.00\%                                                             \\
\multicolumn{1}{|l|}{8}       & \multicolumn{1}{l|}{M1 core}                & \multicolumn{1}{l|}{1340} & \multicolumn{1}{l|}{25/25}     & 2/2      & 1/1      & 33.54\%                                                            \\
\multicolumn{1}{|l|}{9}       & \multicolumn{1}{l|}{Reed Solomon Decoder}   & \multicolumn{1}{l|}{2869} & \multicolumn{1}{l|}{124/125}   & 13/15    & 0/0      & 45.75\%                                                            \\
\multicolumn{1}{|l|}{10}      & \multicolumn{1}{l|}{AES}                    & \multicolumn{1}{l|}{3968} & \multicolumn{1}{l|}{0/0}       & 0/0      & 0/0      & 0.00\%                                                             \\ 
\multicolumn{1}{|l|}{11}      & \multicolumn{1}{l|}{Canny edge detector}    & \multicolumn{1}{l|}{4706} & \multicolumn{1}{l|}{17/18}     & 19/20    & 0/0      & 32.41\%                                                            \\ \hline
                              &                                             &                           &                                &          & Average  & 27.04\%                                                            \\ \cline{6-7} 

\end{tabular}
\end{adjustbox}
\end{table}

\begin{table}[!htb]
\centering
\caption{Gate Coverage}
\label{GC_COMP}
\begin{adjustbox}{width=1\linewidth}
\begin{tabular}{cl|c|c|c|}
\hline
\multicolumn{1}{|l|}{Sl. No.} & Benchmark              & Gates   & \begin{tabular}[c]{@{}l@{}}Module\\ Coverage (\%)\end{tabular} & \begin{tabular}[c]{@{}l@{}}Known Component\\ Coverage (\%)\end{tabular} \\ \hline
\multicolumn{1}{|l|}{1}       & Simple 8-bit processor & 133     & 60.20\%                                                        & 60.20\%                                                                 \\
\multicolumn{1}{|l|}{2}       & Hilbert transformer    & 817     & 100.00\%                                                       & 100.00\%                                                                \\
\multicolumn{1}{|l|}{3}       & UART                   & 898     & 54.23\%                                                        & 54.23\%                                                                 \\
\multicolumn{1}{|l|}{4}       & Cordic polar2rect      & 1602    & 100.00\%                                                       & 48.22\%                                                                 \\
\multicolumn{1}{|l|}{5}       & Cordic rect2polar      & 2331    & 97.17\%                                                        & 40.70\%                                                                 \\
\multicolumn{1}{|l|}{6}       & OpenFPU                & 2466    & 91.81\%                                                        & 63.34\%                                                                 \\
\multicolumn{1}{|l|}{7}       & LXP32 processor        & 3325    & 75.78\%                                                        & 34.19\%                                                                 \\
\multicolumn{1}{|l|}{8}       & M1 core                & 4284    & 72.03\%                                                        & 52.74\%                                                                 \\
\multicolumn{1}{|l|}{9}       & Reed Solomon decoder   & 6754    & 54.78\%                                                        & 54.78\%                                                                 \\
\multicolumn{1}{|l|}{10}      & 128-bit AES core       & 7676    & 39.24\%                                                        & 39.24\%                                                                 \\ 
\multicolumn{1}{|l|}{11}      & Canny edge detector    & 12588   & 81.61\%                                                        & 72.65\%                                                                 \\ \hline
                              &                        & Average & 75.17\%                                                        & 56.12\%                                                                 \\ \cline{3-5} 

\end{tabular}
\end{adjustbox}
\end{table}

%% file: Future_work.tex
\section{Conclusions and Future work}

This paper discussed an automated tool-chain to detect word-level operations and datapath components in flattened LUT-level netlists extracted from bitstreams and express the result using a high-level model. Carry-chain analysis and k-cut analysis are used for identifying combinational modules. Grouping techniques and existing graph isomorphism techniques are used to identify sequential components. The paper also highlights the importance of analyzing carry-chains for reverse engineering FPGAs. We intend to extend the carry-chain analysis to identify multipliers, dividers and counters. Identification of modular boundary of ALUs in real world designs is another problem that requires a robust solution.

%% file: Acknowledgment.tex
\section*{Acknowledgement}

This work was supported in part by the National Science Foundation under IUCRC-1916762, and CHEST (Center for Hardware Embedded System Security and Trust) industry funding.